\documentclass[twocolumn,aps]{revtex4}
\usepackage{graphicx}
\usepackage{amsmath,amssymb}

\begin{document}
\title{Resonant tunneling controlled by laser and constant electric fields}
\author{J. Z. Kami\'{n}ski\thanks{e-mail: jkam@fuw.edu.pl} and E. Saczuk}
\affiliation{Institute of Theoretical Physics, Faculty of Physics,
University of Warsaw, Ho\.{z}a 69, 00-681 Warszawa, Poland
}
\date{\today}

\begin{abstract}
We develop the concept of scattering matrix and we use it to perform 
stable numerical calculations of resonant tunneling of electrons through a 
multiple potential barrier in a semiconductor heterostructure. Electrons move 
in two external nonperturbative electric fields: constant and oscillating in time. 
We apply our algorithm for 
different strengths and spatial configurations of the fields.   
\bigskip

\end{abstract}
\maketitle

\section{Introduction}
The aim of this paper is to present a numerically stable algorithm 
for investigations of nonrelativistic quantum processes 
occurring in arbitrary space-dependent
scalar potential and a time- and space-dependent vector potential. 
Vector potential is
periodic in time and describes a laser field. Such conditions are met
 for example in semiconductor nanostructures
\cite{TG1975,B1987,FG1989,K1993,MBG2009,L2007,L2006,LA2005}
(like quantum wires or wells), carbon nanotubes
\cite{KDB2009,HR2006} or in surface physics 
\cite{FAGAS2009,SMAMK2008,BM2008,FGM2007,DKF2006}.
To make our presentation as clear as possible we shall restrict
ourself to the one-space-dimensional case, although extension of this
algorithm to two and three-space dimensional systems, 
also with magnetic field accounted for,  is possible 
(see, e.g. \cite{FKS2005})
. We shall apply our method
to investigation of the tunneling process and its dependence on
relative phases of multi-chromatic laser pulses (multi-color processes
have been considered for instance in \cite{DKF2006,A2009}).

Multiple barrier, field-assisted resonant tunneling is an interesting problem
because it provides an insight into the physics of nanostructure
quantum systems and because it is a fundamental effect to use in a wide variety
of technological applications. As to the latter, it is enough to mention
all sorts of detectors and generators of microwave radiation based on
double barrier structures with external electric field added;
for more examples, see~\cite{R2003,BB02,D1998,TE73,R86,WV91}.
Here we analyze resonant tunneling through semiconductor
structures in the presence of both oscillating and constant in time external
electric fields. The fields are assumed to be nonperturbative.
We assume that single-particle
states of electrons in heterostructures are well approximated by the 
so-called envelope function \cite{D1998,WV91}.
Effects of sharp interfaces between different 
semiconductors are accounted for by boundary
conditions satisfied by the envelope wavefunction, i.e., by the
continuity of both the envelope wavefunction and the probability
current at the interfaces (see, for example, 
\cite{SW02,LL92,KE99,ML01,SK03,K90,FK97,LLA01}).
Scalar potential $V(x)$ is assumed to be constant in time but it can be
of any shape. The same conditions hold 
for space-dependent effective mass $m(x)$.
Vector potential $A(x,t)$ describes a laser field and thus it 
is space-dependent and oscillates in time.
In our approach to numerical computations, one-dimensional
space is sliced into small intervals where $m(x)$, $V(x)$ and
$A(x,t)$ are space-independent. For arbitrary
space-dependent functions $m(x)$, $V(x)$ and $A(x,t)$ such a
procedure is justified provided that the widths of these intervals
are sufficiently small. We develop below a general numerical scheme
which permits to evaluate transition and reflection probabilities
for electrons moving in the system described above.

This paper is organized as follows. 
In Sec. II the most general solution of the
Schr\"odinger equation is introduced. 
The transfer-matrix method and matching
conditions are analyzed in Sec. III, 
whereas reflection and transition probabilities
are introduced in Sec. IV. These probabilities 
must sum up to 1, which puts a very
strong check for the accuracy of numerical 
calculations. The most important part of
this paper, i.e. the concept of the 
scattering-matrix method, is discussed in the next
section, where it is shown why the scattering-matrix algorithm has to be introduced, instead of a much
simpler transfer-matrix algorithm. Numerical illustrations of the applicability of this algorithm are presented in Sections VI and VII, 
and are followed by short conclusions.

\section{Solution of the Schr\"odinger equation}

\begin{figure}[t]
\begin{center}
\includegraphics[width=7.0cm,height=6.0cm]{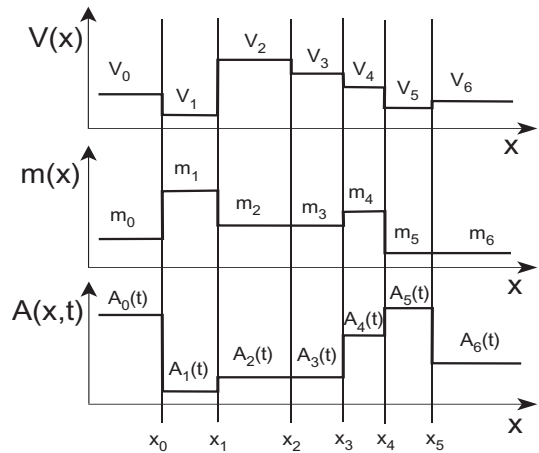}
\end{center}
\caption{Generic shapes of space-dependent superlattice potential
$V(x)$, effective mass $m(x)$, and oscillating in time 
laser field $A(x,t)$. } \label{f1}
\end{figure}

Let us start with one-dimensional Schr\"odinger
equation of the form \cite{LL92,D1998},
\begin{align}
\mathrm{i}\partial_{t}\psi(x,t)=&\Bigl [ \frac{1}{2} \Bigl
(\frac{1}{\mathrm{i}}\partial_{x}-eA(x,t)\Bigr
)\frac{1}{m(x)}\Bigl (\frac{1}{\mathrm{i}}
\partial_{x}-eA(x,t)\Bigr )\nonumber \\
+&V(x)\Bigr ]
\psi(x,t).
\label{sec_1_1}
\end{align}
Space-dependent mass $m(x)$, scalar potential $V(x)$ and
 vector potential $A(x,t)$ are spatially constant in finite
intervals. Their values in any interval $(x_{i-1},x_{i})$
will be denoted as $m_{i}$, $V_{i}$ and
$A_{i}(t)$. An example of such a structure is presented in Fig.
\ref{f1}. We require also that the function $A(x,t)$ is periodic
in time, that is
\begin{equation}
A(x,t+T)=A(x,t),
\label{sec_1_2}
\end{equation}
where $T=2\pi/\omega$ and $\omega$ is the frequency of the
oscillating in time electric field.
Defining in a standard way the probability density
$\rho(x,t)$,
\begin{equation}
\rho(x,t)={\arrowvert\psi(x,t)\arrowvert}^{2},
\label{sec_1_3}
\end{equation}
and the probability current $j(x,t)$,
\begin{eqnarray}
j(x,t)&=&\frac{1}{2}{\psi^{\mathrm{*}}(x,t)}\frac{1}{m(x)} \Bigl
(\frac{1}{\mathrm{i}}\partial_{x}-eA(x,t)\Bigr )\psi(x,t) \\
&+&\frac{1}{2}\psi(x,t)\frac{1}{m(x)} \Bigl [ \Bigl
(\frac{1}{\mathrm{i}}\partial_{x}-eA(x,t)\Bigr ) \psi(x,t)\Bigr
]^{\mathrm{*}}, \nonumber \label{sec_1_4}
\end{eqnarray}
we show  using Eq. (\ref{sec_1_1}) that the conservation 
of probability condition  is satisfied. Indeed,
assuming the above definitions,
we get the continuity equation,
\begin{equation}
\partial_t \rho(x,t)+\partial_x j(x,t)=0.
\label{sec_1_4a}
\end{equation}
Space dependence of mass in Eq. (\ref{sec_1_1}) 
forces one to impose non-standard continuity
conditions on any solution of this equation. It is now
the wavefunction $\psi(x,t)$ and the quantity
\begin{equation}
\frac{1}{m(x)}\Bigl (\frac{1}{\mathrm{i}}\partial_{x}-eA(x,t)\Bigr
)\psi(x,t) \label{sec_1_5}
\end{equation}
that have  to be continuous at points of discontinuity of mass
$m(x)$ and both potentials $V(x)$ and $A(x,t)$ \cite{LL92,KE99,ML01,SK03}.
Before passing to a general solution $\psi(x,t)$ of Eq. (\ref{sec_1_1})
in any given interval $(x_{i-1},x_{i})$, which we shall denote as
$\psi_{i}(x,t)$, let us note that due to time periodicity of the
Hamiltonian, $\psi_{i}(x,t)$ can be chosen such that the Floquet
condition,
\begin{equation}
\psi_{i}(x,t+T)={\mathrm{e}}^{-\mathrm{i}ET}\psi_{i}(x,t),
\label{sec_1_6}
\end{equation}
is satisfied,
where $E$ is the so-called quasienergy.
A general solution $\psi_{i}(x,t)$ of Eq. (\ref{sec_1_1}) in any
interval $(x_{i-1}, x_{i})$ takes then  the following form \cite{K90,FK97},
\begin{eqnarray}
\psi_{i}(x,t)&=&\sum_{M=-\infty}^{\infty} \exp{\bigl
(-\mathrm{i}(E+M\omega)t\bigr )} \sum_{\sigma=\pm}^{}\label{sec_1_7} \\
 &\times&\sum_{N=-\infty}^{\infty}C_{iN}^{\sigma}\mathcal{B}_{M-N}(\sigma
p_{iN})\exp{(\mathrm{i}\sigma p_{iN}x)},
\nonumber
\end{eqnarray}
where $C_{iN}^{\sigma}$ are arbitrary complex numbers to be determined and
\begin{equation}
p_{iN}=\sqrt{2m_{i}(E+N\omega-V_{i}-U_{i})},
\label{sec_1_8}
\end{equation}
with $U_i=e^2\langle A_i^2(t)\rangle/2m_i$ being the ponderomotive energy,
where $\langle A_i^2(t)\rangle$ means the time-average of
$A_i^2(t)$ over the laser-field oscillation.
Components for which $p_{iN}$ are purely imaginary are called
closed channels. These channels are not observed for a particle in
initial or  final states, but they have to be taken into account
in order to satisfy the unitary condition of the time evolution.
In a general case, the $\mathcal{B}_{M-N}(\sigma p_{iN})$ functions
are components of the following Fourier expansion,
\begin{equation}
\exp{\bigl (\mathrm{i}\Phi_{iN}^{\sigma}(t)\bigr
)}=\sum_{M=-\infty}^{\infty}\exp{(-\mathrm{i}M\omega
t)}\mathcal{B}_{M-N}(\sigma p_{iN})
\end{equation}
 provided that the vector potential $A(x,t)$
is periodic in time.
Functions $\Phi_{iN}^{\sigma}(t)$ are defined as follows:
\begin{equation}
\Phi_{iN}^{\sigma}(t)=
\int_{0}^{t}\Bigl [
\frac{\sigma e }{m_{i}}A_{i}(t)p_{iN}-\frac{e^2}{2m_{i}}\bigl (
A_{i}^2(t)-\langle A_{i}^2(t)\rangle\bigr )
\Bigr ]\mathrm{d} t.
\end{equation}
It is easily seen from the above equation that the 
$\mathcal{B}_{M-N}(\sigma p_{iN})$ functions
depend on the form of the vector potential $A(x,t)$, 
that is on the laser field applied.

\section{Matching conditions and transfer matrix}

Continuity conditions discussed above and applied to a general
solution (\ref{sec_1_7}) of the Schr\"odinger equation
(\ref{sec_1_1}) lead to an infinite chain of equations connecting
constants $C_{iN}^{\sigma}$ in the neighboring domains. These
matching conditions can be written in the matrix form,
\begin{equation}
B(i-1,x_{i-1})C_{i-1}=B(i,x_{i-1})C_{i},
\label{sec_2_1}
\end{equation}
where $C_{iN}^{\pm}=[C_{i}^{\pm}]_{N}$ are the components of the
columns $C_{i}^{\pm}$. The matrices $B(i,x)$ and $C_{i}$ are
defined as follows,
\begin{eqnarray}
B(i,x) =
\left( \begin{array}{cc}
B^{+}(i,x) & B^{-}(i,x) \\
{B'}^{+}(i,x) & {B'}^{-}(i,x)
\end{array} \right),
\,
C_{i} =
\left( \begin{array}{c}
C_{i}^{+} \\
C_{i}^{-}
\end{array} \right).
\label{sec_2_2}
\end{eqnarray}
The elements of $B(i,x)$ can be computed in the following way.

For an arbitrary function $A(x,t)$, periodic in time with the period $T$,
\begin{equation}
A(x,t)=A(x,t+T)
\label{period}
\end{equation}
we have
\begin{equation}
A(x,t)=\sum_{n=-\infty}^{\infty}b_{n}(x)\exp{(-\mathrm{i}n\omega t)},
\label{sec_2_2_a}
\end{equation}
where $\omega=2\pi/T$.
In the interval $(x_{i-1}, x_{i})$  coefficients $b_{n}(x)$ assume 
constant values, which we shall denote as
$b_{i,n}$.
Using the condition of the continuity of the wavefunction $\psi_{i}(x,t)$ 
at the point $x_{i-1}$,
we compute the elements of the matrices
$B^{+}$ and $B^{-}$,
\begin{eqnarray}
{B}^{\pm}(i,x)_{M,N} & = & \mathcal{B}_{M-N} (\pm p_{i,N})
\exp{(\pm \mathrm{i}p_{i,N}x)}.
\label{sec_2_5}
\end{eqnarray}
On the other hand elements of the $B'$ matrix  can be evaluated by
substituting a general solution (\ref{sec_1_7}) 
to the expression (\ref{sec_1_5})
and applying the continuity condition to it at
$x_{i-1}$.  After some algebraic manipulations we 
obtain the following equation,
\begin{eqnarray}
&&\frac{1}{m_{i-1}}\Bigl (\sum_{M=-\infty}^{\infty}\exp{\bigl
(-\mathrm{i}(E+M\omega)t\bigr )}\nonumber \\
&\times&\sum_{\sigma=\pm1}^{}\sum_{N=-\infty}^{\infty}
C_{i-1,N}^{\sigma}{\mathcal{B}}_{M-N}(\sigma p_{i-1,N})
\sigma p_{i-1,N}\nonumber \\
&\times&\exp{(\mathrm{i}\sigma
p_{i-1,N}x_{i-1})} - \sum_{M=-\infty}^{\infty}\exp{\bigl
(-\mathrm{i}(E+M\omega)t\bigr)}\nonumber \\
&\times&\sum_{\sigma=\pm1}^{}\sum_{N,n=-\infty}^{\infty}
eb_{i-1,n}C_{i-1,N}^{\sigma}
\mathcal{B}_{M-N-n}(\sigma p_{i-1,N})\nonumber \\
&\times&\exp{(\mathrm{i}\sigma p_{i-1,N}x_{i-1})}\Bigr ) \nonumber \\
&=& \frac{1}{m_{i}}\Bigl (\sum_{M=-\infty}^{\infty}\exp{\bigl
(-\mathrm{i}(E+M\omega)t\bigr)}\nonumber \\
&\times&\sum_{\sigma=\pm1}^{}\sum_{N=-\infty}^{\infty}
C_{i,N}^{\sigma}\mathcal{B}_{M-N}
(\sigma p_{i,N})\sigma p_{i,N}\nonumber \\
&&\exp{(\mathrm{i}\sigma
p_{i,N}x_{i-1})}- \sum_{M=-\infty}^{\infty}\exp{\bigl
(-\mathrm{i}(E+M\omega)t\bigr)}\nonumber \\
&\times&\sum_{\sigma=\pm1}^{}\sum_{N,n=-\infty}^{\infty}
eb_{i,n}C_{i,N}^{\sigma}\mathcal{B}_{M-N-n}(\sigma p_{i,N})\nonumber \\
&\times&\exp{(\mathrm{i}\sigma p_{i,N}x_{i-1})}\Bigr ).
\label{sec_2_8}
\end{eqnarray}
Suspending the summation over $M$ on both sides of the above equation,
we finally get the expression for the $B'$-matrices,
\begin{eqnarray}
{B'}^{\pm}(i,x)_{M,N} & = & \pm \frac{1}{m_{i}}\mathcal{B}_{M-N}
(p_{i,N})p_{i,N}\exp{(\pm \mathrm{i}p_{i,N}x)} \nonumber \\
& - & \frac{1}{m_{i}}\sum_{n=-\infty}^{\infty}eb_{i,n}
\mathcal{B}_{M-N-n} (\pm p_{i,N})\nonumber \\
&\times&\exp{(\pm \mathrm{i}p_{i,N}x)}.
\label{sec_2_10}
\end{eqnarray}
In this way we obtain a set of equations for vectors $C_{i}$,
\begin{equation}
C_{i} = B_{i} C_{i-1},
\label{sec_2_11}
\end{equation}
where
\begin{equation}
B_{i} = [B(i,x_{i-1})]^{-1} B(i-1,x_{i-1}).
\label{sec_2_12}
\end{equation}
These relations allow to connect a solution in a
given domain $x_{i-1}<x<x_{i}$ with an analogous solution in any
other domain $x_{j-1}<x<x_{j}$,
\begin{equation}
C_{j} = B_{j} B_{j-1} ,\ldots, B_{i+1} C_{i} = \mathcal{T}_{ji} C_{i},
\label{sec_2_7a}
\end{equation}
where $\mathcal{T}_{ji}$ is the so-called transfer 
matrix \cite{TE73,K90,ML01,LLA01,LM04}.

\section{Reflection and transition probabilities}

It is clear now that on the basis of Eq.(\ref{sec_2_7a}) we can
connect solutions in the boundary domains $(-\infty,x_{0})$ and
$(x_{L-1},\infty)$. Values of mass $m(x)$, scalar
potential $V(x)$ and vector potential $A(x,t)$ in these
domains will be denoted as $m_{0}$, $V_{0}$, $A_{0}(t)$ and
$m_{L}$, $V_{L}$, $A_{L}(t)$, respectively. We can then write down
 solutions of (\ref{sec_1_1}) for each of these domains. These
solutions represent incident ($\psi_{\mathrm{inc}}$), reflected
($\psi_{\mathrm{ref}}$) and transmitted ($\psi_{\mathrm{tr}}$) waves, and
take the following form,
\begin{eqnarray}
\psi_{\mathrm{inc}}(x,t)&=&\sum_{M=-\infty}^{\infty}
   \exp{(-\mathrm{i}Et)}\exp{(-\mathrm{i}M\omega t)}\nonumber \\
   &\times&\mathcal{B}_{M}(p_{0})
   \exp(\mathrm{i}p_{0}x),
\label{unit_1}
\end{eqnarray}
\begin{eqnarray}
\psi_{\mathrm{ref}}(x,t)&=&\sum_{N,M=-\infty}^{\infty}
C_{0,N}^{-}\exp{(-\mathrm{i}Et)}
   \exp{(-\mathrm{i}M\omega
   t)}\nonumber \\
   &\times&\mathcal{B}_{M-N}(-p_{N})\exp{(-\mathrm{i}p_{N}x)},
\label{unit_2}
\end{eqnarray}
\begin{eqnarray}
\psi_{\mathrm{tr}}(x,t)&=&\sum_{M=-\infty}^{\infty}C_{L,N}^{+}
\exp{(-\mathrm{i}Et)}
 \exp{(-\mathrm{i}M\omega t)}\nonumber \\
 &\times&\mathcal{B}_{M-N}(q_{N}) \exp{(\mathrm{i}q_{N}x)},
\label{unit_3}
\end{eqnarray}
where
\begin{eqnarray}
p_{N}&=&\sqrt{2m_{0}(E+N\omega-V_{0}-U_{0})}, \nonumber \\
q_{N}&=&\sqrt{2m_{L}(E+N\omega-V_{L}-U_{L})}.
\label{unit_4}
\end{eqnarray}
Constants $C_{0,N}^{-}$ and $C_{L,N}^{+}$ will be denoted from now
on as $\textsc{R}_{N}$ and $\textsc{T}_{N}$, respectively. Using 
continuity conditions for functions defined above, we get the
probability conservation equation for reflection and transition
amplitudes, $\textsc{R}_{N}$ and $\textsc{T}_{N}$,
\begin{equation}
   \sum_{N\geqslant N_{\mathrm{ref}}}^{}
   \frac{p_{N}}{p_{0}}{\arrowvert \textsc{R}_{N}\arrowvert}^{2}
   +\sum_{N\geqslant N_{\mathrm{tr}}}^{}\frac{m_{0}q_{N}}{m_{L}p_{0}}
   {\arrowvert \textsc{T}_{N}\arrowvert}^{2}=1,
\label{unit_5}
\end{equation}
where summations are over such $N$ for which $p_N$ and $q_N$ are real,
i.e., over the open channels.
This equation permits us to interpret
\begin{equation}
P_{\mathrm{R}}(N)=\frac{p_{N}}{p_{0}}{\arrowvert 
\mathrm{R}_{N}\arrowvert}^{2}
\label{unit_6}
\end{equation}
and
\begin{equation}
P_{\mathrm{T}}(N)=\frac{m_{0}q_{N}}{m_{L}p_{0}}{\arrowvert \mathrm{T}_{N}
\arrowvert}^{2}
\label{unit_7}
\end{equation}
as reflection and transition probabilities for a tunneling
process in which absorption ($N>0$) or emission ($N<0$) of
energy $N\omega$ by electrons occurred \cite{K90,SK03}. In case of a
monochromatic laser field this process can be interpreted as 
absorption or emission of $N$ photons from the laser field.

The unitary condition (\ref{unit_5}) can be also interpreted as the conservation of
electric charge. To this end, let us define the quantities proportional to the
density of electric currents,
\begin{eqnarray}
J_{\mathrm{inc}}&=&\frac{p_0}{m_0}, \label{currinc}\\
J_{\mathrm{ref}}&=&\sum_{N\geqslant N_{\mathrm{ref}}}
\frac{p_N}{m_0}{\arrowvert \mathrm{R}_{N}\arrowvert}^{2}, \label{currref}\\
J_{\mathrm{tr}}&=&\sum_{N\geqslant N_{\mathrm{tr}}}
\frac{q_N}{m_L}{\arrowvert \mathrm{T}_{N}\arrowvert}^{2}. \label{currtr}
\end{eqnarray}
Then Eq. (\ref{unit_5}) adopts the form of the first Kirchhoff low,
\begin{equation}
J_{\mathrm{inc}}=J_{\mathrm{ref}}+J_{\mathrm{tr}}. \label{current}
\end{equation}

Using (\ref{sec_2_7a}) we can calculate constants $C_{0,N}^{-}=\mathrm{R}_{N}$
and $C_{L,N}^{+}=\mathrm{T}_{N}$ appearing in equations (\ref{unit_1}) -
(\ref{unit_3}). Indeed, since
\begin{equation}
C_{L}=\mathcal{T}C_{0},
\label{unit_8}
\end{equation}
where transfer matrix $\mathcal{T}=\mathcal{T}_{L0}$, and
because $\mathcal{T}$, $C_{0}$
and $C_{L}$ adopt the following block forms,
\begin{eqnarray}
\mathcal{T}=
\left( \begin{array}{cc}
\mathcal{T}^{++} & \mathcal{T}^{+-} \\
\mathcal{T}^{-+} & \mathcal{T}^{--}
\end{array} \right),
C_{0} =
\left( \begin{array}{c}
C_{0}^{+} \\
\mathrm{R}
\end{array} \right),
C_{L}=
\left( \begin{array}{c}
\mathrm{T} \\
0
\end{array} \right),
\label{unit_9}
\end{eqnarray}
we arrive at
\begin{eqnarray}
\textsc{T}&=&\mathcal{T}^{++}C_{0}^{+}+\mathcal{T}^{+-}\textsc{R}, \nonumber \\
0&=&\mathcal{T}^{-+}C_{0}^{+}+\mathcal{T}^{--}\textsc{R}, \label{unit_11}
\end{eqnarray}
where $\textsc{R}$ and $\textsc{T}$ denote the columns of $\textsc{R}_{N}$ i
$\textsc{T}_{N}$, and $[C_{0}^{+}]_{N}=\delta_{0,N}$. Thus, after some algebraic
manipulations, we have,
\begin{eqnarray}
\textsc{R}&=&-(\mathcal{T}^{--})^{-1}\mathcal{T}^{-+}C_{0}^{+}. \nonumber \\
\textsc{T}&=&\bigl (\mathcal{T}^{++}-\mathcal{T}^{+-}
    (\mathcal{T}^{--})^{-1}\mathcal{T}^{-+}\bigr )C_{0}^{+},
\label{unit_12}
\end{eqnarray}
which allows us to determine the quantities $\textsc{R}_{N}$ and
$\textsc{T}_{N}$ for a given transfer matrix $\mathcal{T}$. For
open channels, these quantities are the amplitudes of 
reflection ($\textsc{R}_{N}$) and transition ($\textsc{T}_{N}$)
probabilities, from which one can compute reflection and
transition probabilities using equations (\ref{unit_6}) and
(\ref{unit_7}). In all our numerical illustrations, condition
(\ref{unit_5}) is satisfied with an error smaller than $10^{-14}$.

\section{The scattering matrix}

We note from equations (\ref{sec_2_5}) and (\ref{sec_2_10}) that each of 
the $B_{i}$ matrices that constitute the
transfer matrix $\mathcal{T}_{ji}$
contain elements $\exp(\pm \mathrm{i}p_{i,N}x_{i})$ that depend on the 
$x_{i}$ coordinates at which
the discontinuities appear. For closed channels, that is when the 
$p_{i,N}$ momenta are purely imaginary,
these numbers are real and may assume arbitrary values, very large 
or very small, depending again on the $x_{i}$ coordinates. Number of the
$B_{i}$ matrices is equal to the number of discontinuity points, that is it 
depends on how we divide the
space into short intervals in order to make our potential tractable by 
our algorithm. It may
therefore turn out that in order to compute the transfer matrix 
$\mathcal{T}_{ji}$, we have to multiply a large number
of the $B_{i}$ matrices, each containing both very small and very 
large numbers. It is clear that such a procedure is numerically unstable.
We have to find a way to modify our method of 
calculations in order to compute the elements of each $B_{i}$
matrix at the same point $x=0$ independently of where the `real' 
$x_{i}$ is. This would eliminate "dangerous"
$\exp(\pm \mathrm{i}p_{i,N}x_{i})$ elements (turning them to $1$),
however at the cost of appearing somewhere else.
We shall see later that these `left-overs' of the
shift into $x=0$ appear only as differences $x_{i+1}-x_{i}$ 
and therefore do not cause any harmful side-effects.
We shall see now that such a modification is possible and the price we 
pay for it is worth the effort.

It follows from Eq. (\ref{sec_2_7a}) that in the neighboring
domains, $(x_{i-2},x_{i-1})$ and $(x_{i-1},x_{i})$, we have,
\begin{equation}
C_{i} = \mathcal{T}_{i,i-1} C_{i-1}.
\label{sec_3_1}
\end{equation}
Although the elements of the transfer matrix 
$\mathcal{T}_{i,i-1}$ have been computed
from the continuity conditions at point $x_{i-1}$, one
can compute them at any other point, for example $x=0$. To this end, let us
notice what follows from the form of the solution (\ref{sec_1_7}). 
Translation of the system by a certain distance $\delta$ along the $x$-axis
causes only multiplication of each member of the sum over $N$ in
(\ref{sec_1_7}) by a constant  $\exp{(\mathrm{i}\sigma p_{iN}\delta)}$. These
constants can be included in coefficients $C_{iN}^{\sigma}$. In this way
we get a new set of constants which we shall denote as
$\tilde{C}_{iN}^{\sigma}$,
\begin{equation}
\tilde{C}_{iN}^{\sigma}=\exp{(\mathrm{i}\sigma p_{iN}\delta)} C_{iN}^{\sigma}.
\label{sec_3_2}
\end{equation}
We shall interpret these constants as coefficients in solution
(\ref{sec_1_7}),
given by the continuity conditions at  point 
$x_{i-1}-\delta$. Eq. (\ref{sec_3_2}) written
in the matrix form becomes,
\begin{equation}
\tilde{C}_{i}=\mathcal{P}_{i}(\delta)C_{i}
\label{sec_3_3}
\end{equation}
where
\begin{equation}
\mathcal{P}_{i}(\delta)=
\left( \begin{array}{cc}
 P_{i}^{+}(\delta) & 0 \\
 0 & P_{i}^{-}(\delta)
\end{array} \right),
\label{sec_3_4b}
\end{equation}
and
\begin{equation}
C_{i} =
\left( \begin{array}{c}
C_{i}^{+} \\
C_{i}^{-}
\end{array} \right),
\tilde{C}_{i} =
\left( \begin{array}{c}
\tilde{C}_{i}^{+} \\
\tilde{C}_{i}^{-}
\end{array} \right).
\label{sec_3_4}
\end{equation}
In the equation above $P_{i}^{\sigma}(\delta)$ is a diagonal matrix,
\begin{equation}
[P_{i}^{\sigma}(\delta)]_{NN'}=\delta_{NN'}\exp{(\mathrm{i}\sigma p_{iN}\delta)},
\label{sec_3_5}
\end{equation}
whereas $C_{i}^{\pm}$ and $\tilde{C}_{i}^{\pm}$ are the columns of the constants
 $C_{iN}^{\pm}$ and $\tilde{C}_{iN}^{\pm}$ respectively, that is
$[C_{i}^{\pm}]_{N}=C_{iN}^{\pm}$ and
$[\tilde{C}_{i}^{\pm}]_{N}=C_{iN}^{\pm}$.
It follows from the form of the matrix $\mathcal{P}_{i}(\delta)$ that the
following relations are satisfied:
\begin{equation}
\mathcal{P}_{i}^{-1}(\delta)=\mathcal{P}_{i}(-\delta),
\label{sec_3_6}
\end{equation}
\begin{equation}
\mathcal{P}_{i}(\delta_{1})\mathcal{P}_{i}(\delta_{2})
=\mathcal{P}_{i}(\delta_{1}+\delta_{2}).
\label{sec_3_7}
\end{equation}

\begin{figure}[t]
\begin{center}
\includegraphics[width=7.0cm]{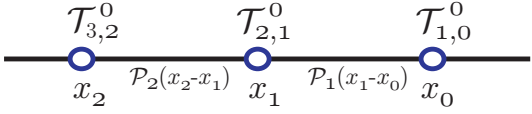}
\end{center}
\caption{Diagrammatic representation of Eq. (\ref{sec_3_13}).
Circles represent points of discontinuity $\{x_j\}$ and matrices 
$\{\mathcal{T}_{j+1,j}^{0}\}$, whereas lines  represent 
`free-propagators' $\{\mathcal{P}_{j+1}(x_{j+1}-x_{j})\}$. 
It is important to notice that all matrices
$\{\mathcal{T}_{j+1,j}^{0}\}$ are calculated at $x=0$, which 
prevents the development of numerical overflows.}
 \label{f2}
\end{figure}

Let us notice also that  translation of the system defined above modifies
the transfer matrix $\mathcal{T}_{i,i-1}$. We have
\begin{eqnarray}
\mathcal{P}_{i}^{-1}\tilde{C}_{i}=C_{i}
&=&\mathcal{T}_{i,i-1}C_{i-1}\nonumber \\
&=&\mathcal{T}_{i,i-1}\mathcal{P}_{i-1}^{-1}(\delta)
\mathcal{P}_{i-1}(\delta)C_{i-1},
\label{sec_3_8}
\end{eqnarray}
thus
\begin{eqnarray}
\tilde{C}_{i}=\mathcal{P}_{i}(\delta)\mathcal{T}_{i,i-1}\mathcal{P}_{i-1}^{-1}
(\delta)\tilde{C}_{i-1},
\label{sec_3_9}
\end{eqnarray}
and we can write it down as
\begin{equation}
\tilde{C}_{i}=\tilde{\mathcal{T}}_{i,i-1}\tilde{C}_{i-1},
\label{sec_3_10}
\end{equation}
where
\begin{equation}
\tilde{\mathcal{T}}_{i,i-1}=\mathcal{P}_{i}(\delta)\mathcal{T}_{i,i-1}
\mathcal{P}_{i-1}^{-1}(\delta).
\label{sec_3_11}
\end{equation}
Matrix elements denoted with the tilde symbol refer to the translated
system.  Using the method defined above and the relation (\ref{sec_2_7a}), we
can connect now the solution in the domain $(-\infty,x_{0})$ with the solution
in any other domain $(x_{i-1},x_{i})$. In this way the elements of the transfer
matrix, which have been computed until now at the points of discontinuity
$x_{0}\ldots x_{i-1}$, are computed now each time at the same point $x=0$. Let us
illustrate this method for a special case of $i=3$
\begin{eqnarray}
C_{3}&=&\mathcal{T}_{3,2}\mathcal{T}_{2,1}\mathcal{T}_{1,0}C_{1}=
\mathcal{P}_{3}^{-1}(x_{2})\mathcal{T}_{3,2}^{0}\mathcal{P}_{2}(x_{2})
\mathcal{P}_{2}^{-1}(x_{1})\nonumber \\
&\times&\mathcal{T}_{2,1}^{0}
\mathcal{P}_{1}(x_{1})\mathcal{P}_{1}^{-1}(x_{0})\mathcal{T}_{1,0}^{0}
\mathcal{P}_{0}(x_{0})C_{0} \nonumber \\
&=&\mathcal{P}_{3}^{-1}(x_{2})\mathcal{T}_{3,2}^{0}
\mathcal{P}_{2}(x_{2}-x_{1})\nonumber \\
&\times&\mathcal{T}_{2,1}^{0}\mathcal{P}_{1}(x_{1}-x_{0})
\mathcal{T}_{1,0}^{0}\mathcal{P}_{0}(x_{0})C_{0}.
\label{sec_3_12}
\end{eqnarray}
Equation (\ref{sec_3_12}) connects  constants $C_{0}$ and $C_{3}$ using the
matrices
$\mathcal{T}_{j,j-1}^{0}$ all computed at $x=0$ independently of $j$, and
 diagonal matrices $\mathcal{P}_{j}(\delta_{j})$, given by the relations
(\ref{sec_3_4b}) and (\ref{sec_3_5}), where $\delta_{j}=(x_{j}-x_{j-1})$.
Edge matrices $\mathcal{P}_{0}(x_{0})$ and $\mathcal{P}_{3}^{-1}(x_{2})$
in the equation (\ref{sec_3_12}) can be omitted 
while computing the transmission and
reflection probability amplitudes since their only role is to multiply the
amplitudes by  phase quotients which disappear while computing the
probabilities. Although these matrices lead to significant modifications of the
closed channels in the domains of $x<x_{0}$ and $x>x_{3}$ in this particular
case, these channels do not influence the reflection and transition 
amplitudes. Transmission and reflection probabilities 
can thus be computed  using a modified transfer matrix,
\begin{equation}
\mathcal{T}_{3,0}^{0}=\mathcal{T}_{3,2}^{0}\mathcal{P}_{2}(x_{2}-x_{1})
\mathcal{T}_{2,1}^{0}\mathcal{P}_{1}(x_{1}-x_{0})\mathcal{T}_{1,0}^{0}.
\label{sec_3_13}
\end{equation}
The matrices $\mathcal{T}_{i,i-1}^{0}$ are equal to the matrices $B_i$
in Eq. (\ref{sec_2_12}) calculated however for $x_{i-1}=0$.
This fact speeds up numerical calculations since now matrix 
$B(i,x=0)$ in Eq. (\ref{sec_2_12}) have to be inverted only once.
Further on we shall omit the superscript $0$ in $\mathcal{T}$ and the tilde over $C$
in order to simplify notation. Diagrammatic representation of the equation above
is shown in Fig. \ref{f2}.

\begin{figure}[t]
\begin{center}
\includegraphics[width=7.0cm]{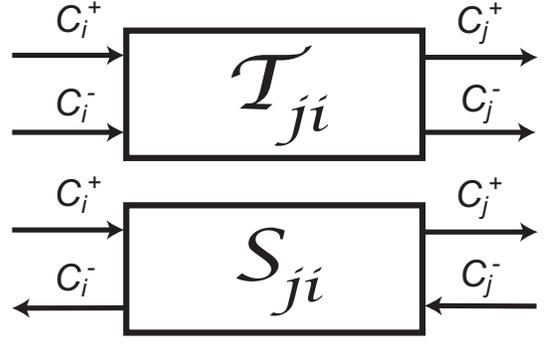}
\end{center}
\caption{Schematic representation of the idea of 
the transfer matrix and the scattering matrix.
For the transfer matrix the incoming channels are 
$C_i^+$ and  $C_i^-$, and the outgoing
channels are $C_j^+$ and  $C_j^-$. For the 
scattering matrix $C_i^+$ and  $C_j^-$ are the incoming
channels with the remaining two considered as the outgoing ones.}
 \label{f3}
\end{figure}

The method presented above is still numerically unstable. The reason for this
instability lies in the existence of large numerical values of elements of
the  $\mathcal{P}_{i}^{-}(\delta)$ matrix for imaginary momenta $p_{iN}$.
In other words, for
\begin{eqnarray}
C_{i}&=&
\left( \begin{array}{c}
C_{i}^{+} \\
C_{i}^{-}
\end{array} \right)=
\mathcal{T}_{i,i-1}C_{i-1}\nonumber \\
&=&
\left( \begin{array}{cc}
\mathcal{T}_{i,i-1}^{++}  & \mathcal{T}_{i,i-1}^{+-}\\
\mathcal{T}_{i,i-1}^{-+}  & \mathcal{T}_{i,i-1}^{--}
\end{array} \right)
\left( \begin{array}{c}
C_{i-1}^{+} \\
C_{i-1}^{-}
\end{array} \right),
\label{sec_3_14}
\end{eqnarray}
the source of  numerical instabilities are matrix elements
$\mathcal{T}_{i,i-1}^{--}$ that contain large numbers.
There is however a chance  for improving the stability,
if only its reverse will be used, 
$\bigl (\mathcal{T}_{i,i-1}^{--}\bigr )^{-1}$.
It appears that it is possible provided 
that in our numerical algorithm only the
so-called scattering matrix will be applied. For this reason
we will show below how to compute the scattering matrix,
$\mathcal{S}_{j,i}$, using only elements of
the transfer matrix, $\mathcal{T}_{j,i}$.
For the transfer matrix $\mathcal{T}_{j,i}$ we have,
\begin{eqnarray}
\mathcal{T}_{j,i}C_{i}=C_{j}=
\left( \begin{array}{c}
C_{j}^{+} \\
C_{j}^{-}
\end{array} \right)=
\left( \begin{array}{cc}
\mathcal{T}_{j,i}^{++}  & \mathcal{T}_{j,i}^{+-}\\
\mathcal{T}_{j,i}^{-+}  & \mathcal{T}_{j,i}^{--}
\end{array} \right)
\left( \begin{array}{c}
C_{i}^{+} \\
C_{i}^{-}
\end{array} \right).
\label{sec_3_15}
\end{eqnarray}
Thus,
\begin{eqnarray}
C_{j}^{+}&=&\mathcal{T}_{j,i}^{++}C_{i}^{+}+\mathcal{T}_{j,i}^{+-}C_{i}^{-},
\nonumber \\
C_{j}^{-}&=&\mathcal{T}_{j,i}^{-+}C_{i}^{+}+\mathcal{T}_{j,i}^{--}C_{i}^{-}.
\label{sec_3_16}
\end{eqnarray}
On the basis of (\ref{sec_3_16}) we now want to compute the elements of the
$\mathcal{S}_{j,i}$ matrix. This matrix 
is supposed to connect the coefficients
$C_{i}^{\pm}$ and $C_{j}^{\pm}$ in the following 
way (for the graphical illustration, see Fig. \ref{f3}),
\begin{eqnarray}
\left( \begin{array}{c}
C_{i}^{-} \\
C_{j}^{+}
\end{array} \right)=
\left( \begin{array}{cc}
\mathcal{S}_{j,i}^{++}  & \mathcal{S}_{j,i}^{+-}\\
\mathcal{S}_{j,i}^{-+}  & \mathcal{S}_{j,i}^{--}
\end{array} \right)
\left( \begin{array}{c}
C_{i}^{+} \\
C_{j}^{-}
\end{array} \right).
\label{sec_3_17}
\end{eqnarray}
Using the set of linear equations (\ref{sec_3_16}), we easily compute the
coefficients
$C_{i}^{-}$ and $C_{j}^{+}$ on the left-hand side of equation
(\ref{sec_3_17}),
as functions of the coefficients $C_{j}^{-}$ and $C_{i}^{+}$.
We get then the following relations,
\begin{eqnarray}
C_{i}^{-}&=&(\mathcal{T}_{j,i}^{--})^{-1}
(C_{j}^{-}-\mathcal{T}_{j,i}^{-+}C_{i}^{+}),
\nonumber \\
C_{j}^{+}&=&\bigl ( \mathcal{T}_{j,i}^{++}-\mathcal{T}_{j,i}^{+-}
(\mathcal{T}_{j,i}^{--})^{-1}
\mathcal{T}_{j,i}^{-+}\bigr )C_{i}^{+}\nonumber \\
&+&\mathcal{T}_{j,i}^{+-}
(\mathcal{T}_{j,i}^{--})^{-1}C_{j}^{-}.
\label{sec_3_18}
\end{eqnarray}
Finally we compute the elements of the matrix $\mathcal{S}_{j,i}$,
\begin{eqnarray}
\mathcal{S}_{j,i}^{++}&=&-(\mathcal{T}_{j,i}^{--})^{-1}
\mathcal{T}_{j,i}^{-+}, \nonumber \\
\mathcal{S}_{j,i}^{+-}&=&(\mathcal{T}_{j,i}^{--})^{-1}, \nonumber \\
\mathcal{S}_{j,i}^{-+}&=&\bigl ( \mathcal{T}_{j,i}^{++}-\mathcal{T}_{j,i}^{+-}
(\mathcal{T}_{j,i}^{--})^{-1}\mathcal{T}_{j,i}^{-+}\bigr ), \nonumber \\
\mathcal{S}_{j,i}^{--}&=&\mathcal{T}_{j,i}^{+-}
(\mathcal{T}_{j,i}^{--})^{-1}.
\label{sec_3_19}
\end{eqnarray}
As expected, the matrix $\mathcal{S}_{j,i}$ contains only  numerically
stable elements $(\mathcal{T}_{j,i}^{--})^{-1}$.

\begin{figure}
\begin{center}
\includegraphics[width=7.0cm]{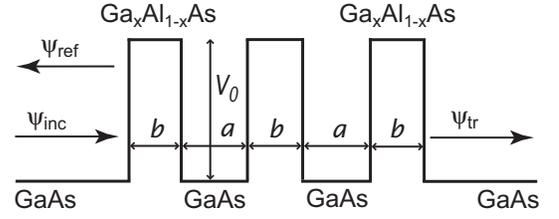}
\end{center}
\caption{Tunneling process considered in this paper. Parameters
for the triple barrier are: $V_0=237$meV,
and the effective masses $m_{\mathrm{GaAs}}=0.0667m_{\mathrm{e}}$ and
$m_{\mathrm{Ga_xAl_{1-x}As}}=0.0918m_{\mathrm{e}}$, 
where $m_{\mathrm{e}}$ is the
electron rest mass. The widths of the barriers $b$ and wells $a$ can change.}
 \label{f4}
\end{figure}

It follows from Eq. (\ref{sec_2_7a}) that the transfer matrix
$\mathcal{T}_{j,i}$ can be written as the product of two transfer matrices,
$\mathcal{T}_{j,k}$ and $\mathcal{T}_{k,i}$ ($i<k<j$),
\begin{eqnarray}
\mathcal{T}_{j,i}=\mathcal{T}_{j,k}\mathcal{T}_{k,i},
\end{eqnarray}
where matrices  $\mathcal{T}_{j,k}$ and $\mathcal{T}_{k,i}$ 
are defined as follows,
\begin{eqnarray}
C_{k}&=&\mathcal{T}_{k,i}C_{i},\nonumber \\
C_{j}&=&\mathcal{T}_{j,k}C_{k}.
\end{eqnarray}
Applying the method presented above, for each of the transfer matrices
$\mathcal{T}_{j,k}$ and $\mathcal{T}_{k,i}$
we can construct a scattering matrix,
$\mathcal{S}_{j,k}$ and $\mathcal{S}_{k,i}$ respectively.
Elements of the scattering matrix $\mathcal{S}_{j,i}$ can
be  computed using only elements of the matrices
$\mathcal{S}_{j,k}$ and $\mathcal{S}_{k,i}$. Using the 
notation above, we obtain the following
expressions for the elements of the $\mathcal{S}_{j,i}$ matrix,
\begin{align}
\mathcal{S}_{j,i}^{++}=&\mathcal{S}_{k,i}^{++}+\mathcal{S}_{k,i}^{+-}
   (1-\mathcal{S}_{j,k}^{++}\mathcal{S}_{k,i}^{--})^{-1}
   \mathcal{S}_{j,k}^{++}\mathcal{S}_{k,i}^{-+},\nonumber \\
\mathcal{S}_{j,i}^{+-}=&\mathcal{S}_{k,i}^{+-}(1-\mathcal{S}_{j,k}^{++}
   \mathcal{S}_{k,i}^{--})^{-1}\mathcal{S}_{j,k}^{+-},\nonumber \\
\mathcal{S}_{j,i}^{-+}=&\mathcal{S}_{j,k}^{-+}(1-\mathcal{S}_{j,k}^{++}
   \mathcal{S}_{k,i}^{--})^{-1}\mathcal{S}_{k,i}^{-+},\nonumber \\
\mathcal{S}_{j,i}^{--}=&\mathcal{S}_{j,k}^{--}
+\mathcal{S}_{j,k}^{-+}\mathcal{S}_{k,i}^{--}
   (1-\mathcal{S}_{j,k}^{++}\mathcal{S}_{k,i}^{--})^{-1}
   \mathcal{S}_{j,k}^{+-}.
\label{sec_3_20}
\end{align}
It is clear from the above that the $\mathcal{S}_{j,i}$ 
matrix is not merely a product of two matrices
$\mathcal{S}_{j,k}$ and $\mathcal{S}_{k,i}$, but rather a complicated
nonlinear composition of them.
It is important however to note that despite its evident complexity, 
such a construction of the scattering matrix
is numerically stable, as opposed to the transfer matrix method which 
fails if a system with a large number of
discontinuity points $x_{i}$ is considered. Stability of such an algorithm
has been proven in our numerical investigations by checking that the condition
(\ref{unit_5}) is satisfied with an error smaller than $10^{-14}$.
Such an accuracy can never be achieved for systems with a large
number of discontinuity points if the transfer matrix is applied.

\begin{figure}
\begin{center}
\includegraphics[width=7.0cm]{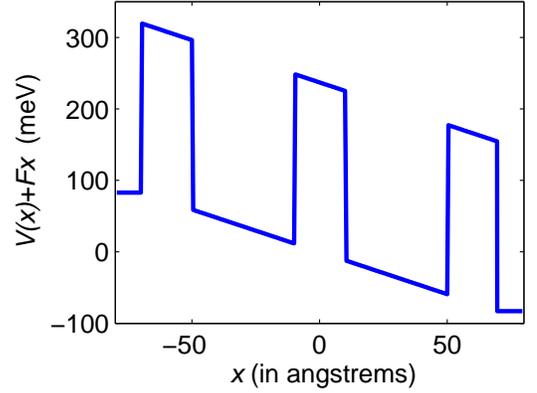}
\end{center}
\caption{(Color online) Plot of the potential $V(x) + Fx$ 
for $a = 40${\AA}, $b = 20${\AA},
and $F =-0.23\times 10^{-4}$(in atomic units). Other 
parameters are the same as in Fig. \ref{f4}.}
 \label{f5}
\end{figure}

\section{Resonant tunneling}

We shall consider now the tunneling phenomenon
through a semiconductor heterostructure presented in
Fig. \ref{f4}. In the beginning, let us assume that  electrons
interact only with a constant electric field. Hence, the
time-independent potential is of the form $V(x) + Fx$, in
which $V(x)$ represents the semiconductor heterostructure
potential (Fig. \ref{f4}) and $F$ is the electric-field strength.
The plot of this potential is presented in Fig. \ref{f5}, where
$a = 40${\AA}, $b = 20${\AA}, and $F =-0.23 \times 10^{-4}$ (in atomic units).
Applying the theory developed above, we calculated
 reflection and transmission probabilities (see Fig. \ref{f5})
discretizing the potential with 15, 141, and 281 points, as
indicated in one of the frames. We observe that, in order
to get convergence, one has to introduce at least one
hundred discontinuity points. There is no noticable difference between
the results obtained for 141 and 281 such points.

\begin{figure}
\begin{center}
\includegraphics[width=7.0cm]{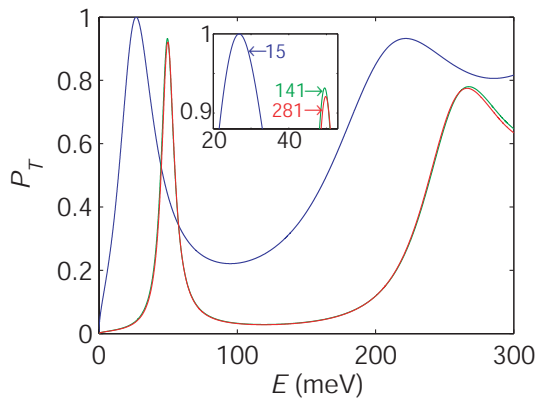}
\end{center}
\caption{(Color online) Transmission probabilities for the potential 
shown in Fig. \ref{f5}. Numbers
in the insert frame (representing enlarged part of the main frame) 
indicate the number of equally spaced 
discontinuity points introduced in our numerical algorithm. We see that
15 points do not give correct results and that the 
convergence is reached with more than 100 points.}
 \label{f6}
\end{figure}

\begin{figure}
\begin{center}
\includegraphics[width=7.0cm]{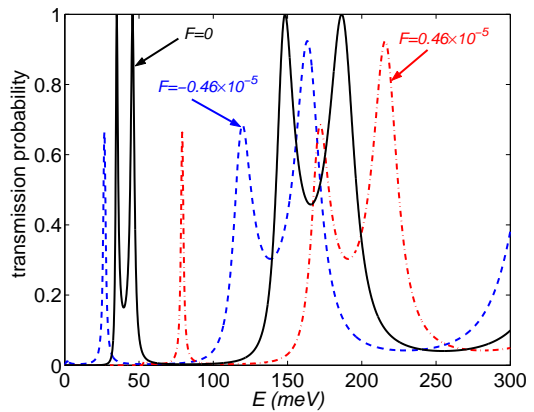}
\end{center}
\caption{(Color online) Transmission probabilities for semiconductor
triple barrier with $a = 70${\AA}, $b = 20${\AA}. Intensity of the
laser field is such that the ratio of ponderomotive energy to
 photon energy is $U_p/\omega = 10^{-4}$ with $\omega= 70$meV and we have
three electric-field strengths (in atomic units), as indicated in the figure.
As expected, transition probability distributions for $\pm F$ [blue
(dash-dash) and red (dash-dot) lines] are shifted
in energy by $|F|(3b+2a)$. Computations were  performed for 221
discretization points.}
 \label{f7}
\end{figure}

\begin{figure}[b]
\begin{center}
\includegraphics[width=7.0cm]{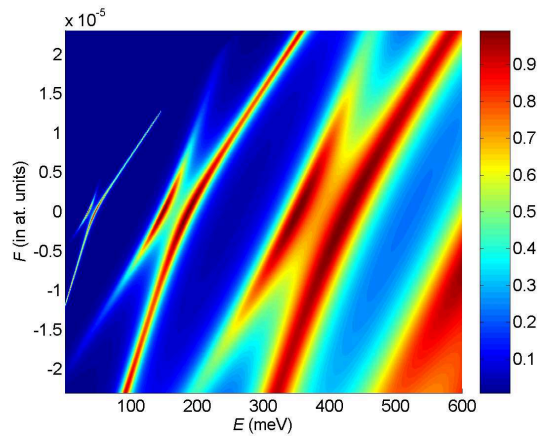}
\end{center}
\caption{(Color online) Color map of the total transmission probability 
in the plane of the incident electron energy $E$ and the electric-field 
strength $F$. As expected, for the vanishing electric field resonances 
in the considered in Fig. \ref{f7} triple barrier structure show up in 
doublets. However, for sufficiently strong electric field the lower-energy 
resonance from a doublet disappears.}
 \label{f8}
\end{figure}

Next, let us analyze transmission of electrons through the triple barrier 
of Fig. \ref{f4} with $a = 70${\AA} and $b = 20${\AA} and with the 221 
discretization points. Now we apply a constant electric field and the 
monochromatic laser field of frequency $\omega= 70\textrm{meV}$ and 
intensity such that its ponderomotive energy divided by the laser photon 
energy equals $10^{-4}$. Without external fields, the resonant energies are 
grouped in doublets in which the lower-energy resonance corresponds to 
the antisymmetric resonance state, and the upper-energy resonance to the symmetric one. 
With the laser field switched on this structure does not change very much 
provided that the frequency is off-resonance with respect to the already 
existing resonance states of the triple barrier, and the intensity is not 
too large, as it is presented in Figs. \ref{f7} and \ref{f8}. The pattern 
changes significantly if a constant electric field is applied. We observe 
that with an increasing strength of the electric field the low-energy 
transmission resonances from a given doublet gradually disappear and we 
are left with a single transmission resonance, which for even stronger 
electric fields also dies out. This means that by proper tunning the 
strength of a constant electric field one can selectively transmit 
electrons of some particular energies.

\section{Phase control of tunneling}

Special features of barrier problems stem from the interaction of waves 
reflected from or transmitted through potential jumps. When the interference 
of reflected waves is in phase, transmission becomes minimal. But when the 
interference of reflected waves is out of phase (i.e., it is destructive) 
the incident wave resonantly penetrates either by tunneling through or 
passing above the barrier structure. If the process occurs in a monochromatic 
laser field the destructive or constructive interferences between reflected 
and transmitted waves are present also for different Fourier components of 
the electron wavefunction. This leads for example to opening or closing gaps 
in the band structure \cite{TG1975,B1987,FG1989,K1993,MBG2009,FK97} or 
formation of multiple-plateau structures in the high-order harmonic spectrum 
\cite{FK96,FKS98}. It gets even more complicated if multichromatic laser 
fields or short laser pulses are applied. In the first case the interference 
discussed above can be controlled by relative phases of harmonics present 
in the multichromatic fields, whereas in the second case the resonance 
transmission can be modified by the carrier-envelope phase.

\begin{figure}
\begin{center}
\includegraphics[width=8.0cm]{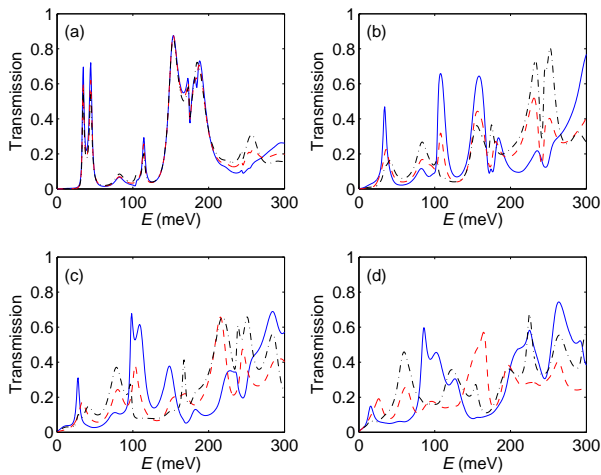}
\end{center}
\caption{(Color online) Total transmission probabilities for the triple 
barrier structure presented in Fig. \ref{f4} with 
$a = 70${\AA} and $b = 20${\AA}, and for
the bichromatic field [Eq. (\ref{bichrom1})] with $\omega=70\textrm{meV}$. The
continuous (blue) line is for $\varphi=0$, 
dash-dash (red) line for $\varphi=\pi/2$,
whereas dash-dot (black) line for $\varphi=\pi$. Frames correspond 
to different laser field intensities characterized 
by the dimensionless parameter
$\xi=|e|E_0(x)/(2\sqrt{\hbar m_{\mathrm{e}}\omega^3})$ ($m_{\mathrm{e}}$ 
is the electron rest mass and $E_0(x)$ is considered to be constant in 
the whole space): (a) $\xi=0.1$, (b) $\xi=0.5$, (c) $\xi=1$ and (d) $\xi=2$.}
 \label{f9}
\end{figure}

\begin{figure}
\begin{center}
\includegraphics[width=8.0cm]{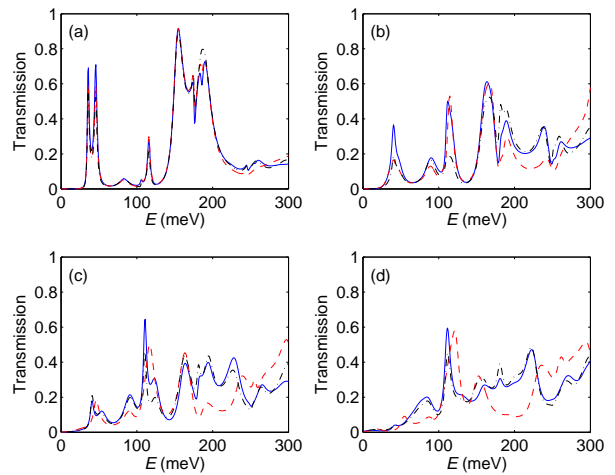}
\end{center}
\caption{(Color online) The same is in Fig. \ref{f9}, 
but with space-dependent
intensity of a laser field. Now, $E_0(x)=E_0$ within 
the triple barrier structure,
$E_0(x)=E_0/2$ within the edge barriers and 0 outside. 
The electric field strength
$E_0$ is determined by a dimensionless parameter 
$\xi=|e|E_0/(2\sqrt{\hbar m_{\mathrm{e}}\omega^3})$, 
with the same numerical values as in Fig. \ref{f9}.}
 \label{f10}
\end{figure}

As an example let us consider a bichromatic laser field. 
Let the electric field be of the form
\begin{equation}
E(x,t)=E_0(x)[\sin(\omega t)-\sin(2\omega t+\varphi)],
\label{bichrom1}
\end{equation}
where $E_0(x)$ is in general a space-dependent 
amplitude of the laser field.
In Figs. \ref{f9} and \ref{f10} the laser-modified 
total transmission probabilities
through a triple-barrier structure are presented for 
three different relative
phases $\varphi$. Fig. \ref{f9} corresponds 
to the situation in which the laser field acts
in the whole space, whereas Fig. \ref{f10} illustrates 
the action of the laser field
concentrated within the structure, hence incident, reflected 
and transmitted electrons are free. Apart from a significant 
dependence of these probabilities on the relative phase we 
observe also that in the second case and for higher intensities 
considered the transmission probabilities are smaller. It is 
because the electrons have to traverse an extra 'potential barrier' 
created by the ponderomotive energy of a laser field in order to 
transmit through barriers. This finding opens up a possibility to 
create tunneling barrier structures by laser fields modulated in 
space. This can be investigated numerically by applying the algorithm 
developed in this paper.  

As a second example of the phase control let us consider transmission 
of electrons through a triple barrier structure in the presence of 
both a constant electric field and a train of very short laser pulses. 
It is well-known from atomic and molecular physics that the ionization 
process can be significantly modified by the so-called carrier-envelope 
phase if a single pulse contains only few oscillations. In order to 
investigate this phenomenon for electron transmission let us assume 
that the train of pulses is build from a single pulse (defined for 
times $0\leqslant t\leqslant T_p$) of the form  
\begin{equation}
E(x,t)=E_0(x)f(t)\sin(\omega t+\varphi)-\Delta_L,
\label{puls1}
\end{equation}
where the envelop function $f(t)$ equals
\begin{equation}
f(t)=\exp\Bigl[-\Bigl(\frac{t-T_p/2}{\sigma_p}\Bigr)^2\Bigr]
\sin^2\Bigl(\frac{\pi t}{T_p}\Bigr),
\label{puls2}
\end{equation}
and the constant in time $\Delta_L$ is chosen such that
\begin{equation}
\int_0^{T_p} E(x,t)\mathrm{d}t=0.
\label{puls3}
\end{equation}
The carrier-envelope phase $\varphi$ can change from 0 to $2\pi$.

\begin{figure}
\begin{center}
\includegraphics[width=8.0cm]{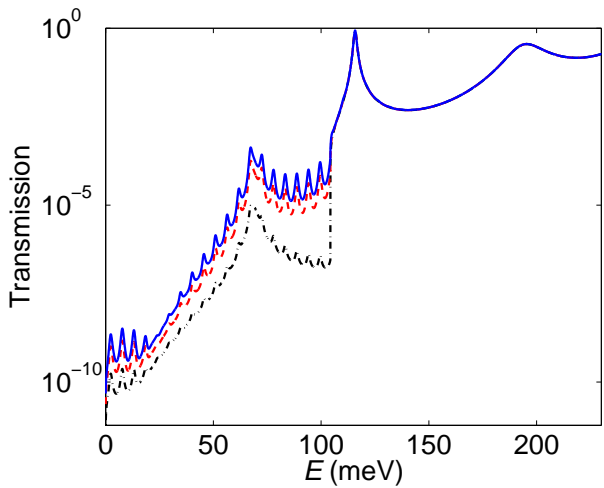}
\end{center}
\caption{(Color online) Total transmission probabilities for the 
triple barrier structure presented in Fig. \ref{f4} with 
$a = 70${\AA}, $b = 20${\AA} and
$V_0=237\textrm{meV}$. The train of laser pulses [Eqs. (\ref{puls1}) 
and (\ref{puls2})] with $\omega=70\textrm{meV}$, $T_p=26\pi/\omega$ 
and $\sigma_p=T_p/140$ (one-cycle pulse) is defined by the 
space-dependent electric field such that
$E_0(x)=E_0$ within the triple barrier structure,
$E_0(x)=E_0/2$ within the edge barriers and 0 outside.
The laser field intensity is characterized by the dimensionless parameter
$\xi=|e|E_0/(2\sqrt{\hbar m_{\mathrm{e}}\omega^3})$ ($m_{\mathrm{e}}$ 
is the electron rest mass), whereas the constant electric field strength 
$F$ is determined by the parameter
$\eta=|e|F(3b+2a)/V_0$. In this illustration $\xi=0.2$ and $\eta=0.1$. 
The continuous (blue) line is for $\varphi=\pi/2$, dash-dash (red) 
line for $\varphi=\pi/4$, whereas dash-dot (black) line for $\varphi=0$.}
 \label{f11}
\end{figure}

In Fig. \ref{f11} we present transmission probabilities for the electrons 
impinging from the right on the triple barrier structure shown in 
Fig. \ref{f5} (however, with different values for $a$ and $F$). 
Without the action of the laser pulse the transmission is forbidden 
for energies smaller than approximately 100\textrm{meV}, whereas for 
larger energies electrons can tunnel resonantly. The presence of the 
laser field modifies these conditions and they get similar 
to those met in the photoemission from solid surfaces or the ionization 
of atoms or molecules. For the latter it is well-known that the carrier-envelope 
phase substantially modifies ionization probabilities \cite{Paulus2001,MPBB2006}. 
The results presented here also confirm this effect for the tunneling 
phenomena. Transmission probabilities (hence, also photocurrents emitted 
from the surface) can then be changed by the carrier-envelope phase even 
by two orders of magnitude.

\section{Conclusions}

As mentioned above, our algorithm is convergent provided that 
a sufficient number of
discretization points is introduced. For systems
considered here, this number should not be smaller
than 100. If the laser field is very weak, this does not
create significant numerical problems, except that
 calculations become longer. However, when the
laser field is sufficiently intense, the algorithm
based on the transfer matrix is unstable. This instability is due
to the existence of closed channels, which introduce
into numerical calculations very small and very large
numbers at the same time. Augmenting precisions significantly slows 
down the calculation and does not diminish the problem.  
We have found that it is
possible to make this algorithm numerically stable by just
applying nonlinear matrix transformations, 
without introducing higher precisions.

Illustrations presented in this paper show that
 tunneling of electrons through 
multi-barrier semiconductor structures can be changed significantly
by applied nonperturbative electric fields:  oscillating in
time or constant. The efficiency of the algorithm presented in this
contribution opens up the possibility of investigating surface phenomena
(like photoemission or high-order harmonic generation) in the presence 
of more realistic laser pulses that gradually decrease within solids
and extend on a mesoscopic scale in vacuum. These problems 
are under investigations.

\acknowledgments This work was supported by the Polish
Ministry of Science and Higher Education (Grant No. N N202 033337).


\begin{thebibliography}{99}

\bibitem{TG1975}
N. Tzoar and J. I. Gersten, Phys. Rev. B {\bf 12}, 1132 (1975).

\bibitem{B1987}
M. R. Beli\'c, Solid State Commun. {\bf 62}, 817 (1987).

\bibitem{FG1989}
F. H. M. Faisal and R. Genieser, Phys. Lett. A {\bf 141}, 297 (1989).

\bibitem{K1993}
J. Z. Kami\'nski, Acta Phys. Pol. A {\bf 83}, 495 (1993).

\bibitem{MBG2009}
D. R. Ma\v{s}ovi\'c, M. R. Beli\'c, and J. I. Gersten, 
Phys. Lett. A {\bf 373}, 3289 (2009).

\bibitem{L2007}
R. P. Lungu, Phys. Scr. {\bf 75}, 206 (2007).

\bibitem{L2006}
R. Lefebvre, Int. J. Quant. Chem. {\bf 106}, 2848 (2006).

\bibitem{LA2005}
R. Lefebvre and O. Atabek, J. Phys. B {\bf 38}, 2133 (2005).

\bibitem{KDB2009}
H. Khosravi, N. Daneshfar, and A. Bahari, Optics Lett. 
{\bf 34}, 1723 (2009).

\bibitem{HR2006}
H. Hsu and L. E. Reichl, Phys. Rev. B {\bf 74}, 115406 (2006); 
{\it ibid.} B {\bf 72}, 155413 (2005).

\bibitem{FAGAS2009}
M. Faraggi, I. Aldazabal, M. S. Gravielle, A. Arnau, 
and V. M. Silkin, J. Opt. Soc. Am. B {\bf 26},
2331 (2009).

\bibitem{SMAMK2008}
G. Saathoff, L. Miaja-Avila, M. Aeschlimann, M. M. Murname, 
and H. C. Kapteyn, Phys. Rev. A {\bf 77}, 022903 (2008).

\bibitem{BM2008}
J. C. Baggesen and L. B. Madsen, Phys. Rev. A {\bf 78}, 032903 (2008).

\bibitem{FGM2007}
M. N. Faraggi, M. S. Gravielle, and D. M. Mitnik, 
Phys. Rev. A {\bf 76}, 012903 (2007).

\bibitem{DKF2006}
P. Dombi, F. Krausz, and G. Farkas, J. Mod. Opt. {\bf 53}, 163 (2006).

\bibitem{FKS2005}
F. H. M. Faisal, J. Z. Kami\'nski, and E. Saczuk, Phys. Rev. A {\bf 72},
023412 (2005); Laser Phys. {\bf 16}, 272 (2006).

\bibitem{A2009}
V. A. Astapenko, Quantum Electron. {\bf 36}, 1131 (2006).

\bibitem{R2003}
M. Razavy, \textit{Quantum Theory of Tunneling} (World Scientific, Singapore, 2003).

\bibitem{BB02}
K. F. Brennan and A. S. Brown, \textit{Theory of Modern Electronic
Semiconductor Devices} (Wiley, New York, 2002).

\bibitem{D1998}
J. H. Davis, \textit{The Physics of Low-dimensional Semiconductors. 
An Introduction} (Cambridge, Cambridge, 1998)

\bibitem{TE73}
R. Tsu and L. Esaki, Appl. Phys. Lett. {\bf 22}, 562 (1973).

\bibitem{R86}
D. K. Roy, \textit{Quantum Mechanical Tunneling and its Applications}
(World Scientific, Singapore, 1986).

\bibitem{WV91}
C. Weisbuch and B. Vinter, \textit{Quantum Semiconductor Structures.
Fundamentals and Applications} (Academic Press, Boston, 1991).

\bibitem{SW02}
W. Sch\"afer and M. Wegener, \textit{Semiconductor Optics and Transport
Phenomena} (Springer, Berlin, 2002).

\bibitem{LL92}
J-M. L\'evy-Leblond, Eur. J. Phys. {\bf 13}, 215 (1992).

\bibitem{KE99}
J. Z. Kami\'nski and F. Ehlotzky, J. Phys. B {\bf 32}, 3193 (1999).

\bibitem{ML01}
N. Moiseyev and R. Lefebvre, Phys. Rev. A {\bf 64}, 052711 (2001).

\bibitem{SK03}
E. Saczuk and J. Z. Kami\'nski, Phys. Stat. Sol. (b) {\bf 240}, 603 (2003).

\bibitem{K90}
J. Z. Kami\'nski, Z. Phys. D {\bf 16}, 153 (1990).

\bibitem{FK97}
F. H. M. Faisal and J. Z. Kami\'nski, Phys. Rev. A {\bf 56}, 748 (1997).

\bibitem{LLA01}
V. Leon, R. Lefebvre and O. Atabek, Phys. Rev. A {\bf 64}, 052105 (2001).

\bibitem{LM04}
R. Lefebvre and N. Moiseyev, Phys. Rev. A {\bf 69}, 062105 (2004).


\bibitem{FK96}
F. H. M. Faisal and J. Z. Kami\'nski, Phys. Rev. A {\bf 54}, R1769 (1996).


\bibitem{FKS98}
F. H. M. Faisal, J. Z. Kami\'nski, and S. S. M. Soliman, Laser Phys. {\bf 8}, 129 (1998).

\bibitem{Paulus2001}
G. G. Paulus, F. Grasbon, H. Walther, P. Villoresi, M. Nisoli, 
S. Stagira, E. Priori, and S. De Silvestri, Nature {\bf 414}, 182 (2001); 
G. G. Paulus, F. Lindner, H. Walther, A. Baltu\v{s}ka, 
E. Goulielmakis, M. Lezius, and F. Krausz, Phys. Rev. Lett. {\bf 91}, 253004 (2003).

\bibitem{MPBB2006}
D. B. Milo\v{s}evi\'c, G. G. Paulus, D. Bauer, and W. Becker, 
J. Phys. B {\bf 39}, R203 (2006).

\end{thebibliography}
\end{document}